\documentclass[twocolumn,prb,showpacs]{revtex4}
\usepackage{graphicx}
 \usepackage{rotating}
\usepackage{amsmath}
\usepackage{amsfonts}
\usepackage{amssymb}
\usepackage{enumerate}
\usepackage{longtable}
\usepackage{dcolumn}
\usepackage{bm}

\begin{document}

\title{Renormalized spin wave excitations in the antiferromagnetic
Heisenberg-Kondo model for heavy fermions}
\author{M. Acquarone$^1$}
\author{C. I. Ventura$^2$}
\affiliation{$^1$IMEM-CNR and Dipartimento di Fisica, Universit\`{a}
di Parma, 43100 Parma, Italy,\email{acquarone@fis.unipr.it} }
\affiliation{$^2$Centro At\'omico Bariloche, 8400 - Bariloche,
Argentina ,  \email{ventura@cab.cnea.gov.ar}}
\date{\textrm{\today}}

\begin{abstract}
  Recent inelastic neutron scattering experiments in CeIn$_{3}$ 
and CePd$_{2}$Si$_{2}$ single crystals, measured spin wave excitations 
at low temperatures. These two heavy fermion compounds 
exhibit antiferromagnetic long-range order, but a strong competition 
between the Ruderman-Kittel-Kasuya-Yosida(RKKY) interaction 
and Kondo effect is evidenced by their nearly equal N\'eel 
and Kondo temperatures. 
Our aim is to show how magnons such as measured in the
antiferromagnetic phase of these Ce compounds, 
can be described with a microscopic Heisenberg-Kondo model    
as introduced by J.R.Iglesias, C.Lacroix and B.Coqblin, 
 used before for studies of the non-magnetic phase.   
The model includes the correlated Ce-$4 f$ electrons   hybridized
with the conduction band, and we consider competing 
RKKY\ (Heisenberg-like $ J_{H} $) and Kondo ($J_{K}$) antiferromagnetic
 couplings. 
Carrying on a series of unitary transformations, we perturbatively 
derive a second-order effective Hamiltonian 
 which, projected onto the antiferromagnetic electron
ground state, describes the spin wave excitations, renormalized by their
interaction with correlated itinerant electrons. We numerically study how the
different parameters of the model influence the renormalization of the
magnons, yielding useful information for the analysis of inelastic neutron
scattering experiments in antiferromagnetic heavy fermion compounds. 
We also compare our results with available experimental data, 
finding good agreement with the spin wave measurements in cubic CeIn$_3$.

\end{abstract}

\maketitle


%


\section{Introduction}

\label{intro}

The description of heavy fermion compounds is challenging due 
to the rich variety of phase diagrams they present,  
and the anomalous physical properties which may be found. 
Among them, appear the Ce and U compounds which exhibit long-range
antiferromagnetism (AF) at low temperatures (for example, among  
antiferromagnetic Ce compounds,  CeRh$_{2}$Si$_{2}$ exhibits 
the highest ordering temperature $T_{N} = 36 K$, 
with local magnetic ordered moments of $1.34 -1.42 \mu_{B}$ per Ce, i.e.
relatively large, compared to the full Ce$^{3+}$ 
 free-ion value: 2.54 $\mu_B$).\cite{rhmom}  
Depending on the particular compound,\cite{radousky} the antiferromagnetism 
takes different forms (magnitude of the local moments 
varies widely: e.g. 0.001 $\mu_{B}$ as in CeRu$_2$Si$_2$ or 0.02 $\mu_{B}$ 
in UPt$_3$, to  1.55 $\mu_{B}$ as in UCu$_5$; 
as does the spin configuration:  with three-, two- or one-dimensional 
AF structures observed).   Antiferromagnetism may also appear
competing or coexisting with superconductivity,\cite{steglich,mathur} 
for which spin fluctuation-mediated pairing mechanisms are
explored.\cite{monod} Non-Fermi liquid behaviour may 
appear,\cite{stewart} and quantum criticality   
has become a subject of intensive study in these compounds, 
both experimentally and theoretically.\cite{mathur,schroeder,kuechler,kitp,
coqblin,coleman}  
The crossover from the antiferromagnetic state to the non-magnetic heavy
fermion state,  which can be tuned by pressure, doping or magnetic field, 
is one of the most interesting problems in strongly correlated $f-$compounds.

The physical properties of these compounds are determined by 
the strongly correlated $f-$electrons present ($4f$ in Ce; 
$5f$ in U) and their hybridization with the conduction band. 
The RKKY indirect exchange interaction between $f-$ local magnetic moments,
 favouring the establishment of long-range magnetic order, 
competes with the screening of these moments by the conduction
 electrons, described by the 
Kondo effect.\cite{kondoeff} 
This competition is the subject 
of the Doniach diagram,\cite{doniach} which compares the variation of the  N\'eel 
and Kondo-impurity temperatures with increasing antiferromagnetic intrasite exchange
coupling $J_K$, between local $f$-moments and conduction electron spins. 
Compounds with similar magnetic ordering temperature $T_N$ and  Kondo
temperature $T_K$, the temperature below which magnetic susceptibility
saturates indicating  coherent Kondo-singlet formation,  
are ideally suited to the study of this RKKY-Kondo competition.
In this regard, experiments on  CeM$_2$Sn$_2$ (M=Ni, Ir, Cu, 
Rh, Pd, and Pt: $4d$ or
$5d$ transition metals)\cite{beyermann}  and CeX$_2$Si$_2$ (X=Au, Pd, Rh, Ru) 
\cite{severing} were undertaken, indicating  that departures 
between theory  and experiments resulted from the use of Kondo 
impurity relations.  
The Kondo-lattice model, instead, consisting of a lattice of local magnetic 
moments coexisting with a conduction band, has proved  appropriate for the description
 of many $4-f$ and $5-f$ materials, in particular 
most Ce (or Yb)  compounds, respectively  corresponding to a configuration 
close to $4f^{1}$ (or  $4f^{13}$), where one $4f$ electron (or hole) 
interacts with the conduction electrons.\cite{coqblin,coqblin1,coqblin2,coqblin3} 
In 1997 a revisited Doniach diagram was introduced, including short-range
antiferromagnetic correlations in the Kondo lattice, in order to improve the
description and, in particular, to account for the observed pressure 
dependence of T$_K$ in CeRh$_2$Si$_2$.\cite{coqblin1}  
The situation is more complex in Uranium compounds,  
where U has a $5f^{n}$ configuration with n=2 or 3, since the $5f$ electrons  
are much less localized than the $4f$ electrons of rare earths. Regarding spin
dynamics, it is not clear that Ce and U compounds are intrinsically
 similar.\cite{radousky} In the 
following we will focus on Ce-compounds, except otherwise specifically stated. 
 
Experimentally, while the magnetic response due to Kondo spin fluctuations in
the paramagnetic state of heavy fermions is well studied,  
relatively little is known about the nature of the 
magnetic excitations in the {\it ordered} phase of Kondo lattices,\cite{radousky} 
on which our present study will focus. Being still an unsolved problem how to 
describe on equal terms both the Kondo effect and
antiferromagnetism,\cite{rech} here we will focus on systems with relatively 
large local moments and study them deep inside the antiferromagnetic phase: 
far from the antiferromagnetic quantum critical point, where spin fluctuations 
would become more relevant.

 A few years ago CePd$_2$Si$_2$\cite{cepd2si2} single crystals  
were studied with inelastic neutron scattering: below the
antiferromagnetic ordering temperature strongly dispersive spin wave
 excitations were found, with an anisotropic damping, 
which coexisted with the Kondo-type spin fluctuations also 
present above $T_N$. At $T=1.5K$ these spin waves were measured along various
 BZ paths: they were found to present an energy gap of 0.83 meV and to 
extend up to almost 3.5 meV. CePd$_2$Si$_2$ has a bcc tetragonal structure, 
 and its antiferromagnetic ground
 state is characterized by propagation vector: $ \vec{q}= (1/2, 1/2, 0)$, 
with ordered moments: $m=0.66 \mu_{B}$, $T_N=8.5K$ and $T_{K}=10K$, and
linear electronic specific heat coefficient $\gamma = 250 mJ/mol K^2$. Under 
pressure application, at 28.6 kbar the system undergoes a transition into a 
superconducting phase with critical temperature of 430 mK.\cite{cein3,llois}   
 More recently, inelastic neutron studies of CeIn$_3$ single crystals  were
performed, with similar results.\cite{cein3}  Well defined spin wave 
excitations with a bandwidth of 2meV and a gap of 1.28 meV  
were found in the antiferromagnetic phase,\cite{cein3} coexisting with Kondo-type spin
fluctuations and crystal-field excitations which also appeared above $T_N = 10K= T_K$.  
CeIn$_3$ crystallizes in a cubic (fcc) structure, 
 with an antiferromagnetic
structure characterized by magnetic propagation vector
$\vec{q}=(1/2, 1/2, 1/2)$, with ordered moments: $m=0.5 \mu_{B}$ and 
$\gamma = 130 mJ/mol K^2$.  Under application of  
pressure, at 26.5 kbar the system undergoes a transition into a 
superconducting phase with critical temperature of 200 mK.\cite{cein3}   

In next section, we will briefly introduce the  microscopic Heisenberg-Kondo 
model proposed by J.R.Iglesias, C.Lacroix and B.Coqblin,\cite{coqblin1} 
to study the non-magnetic phase of heavy fermion AF compounds,
to which we shall add  conduction electron  correlations. 
We will then present our calculation for  the renormalization 
of spin wave excitations 
due to their interaction with the correlated conduction electrons (the Appendix 
complements this section).  
In Section \ref{results}, we will discuss the results of our study,
show how the different parameters of the model influence 
the renormalization of the magnons, 
and compare our results with the available experimental 
data.\cite{cepd2si2,cein3} In Section \ref{conclusions} we summarize and point
out that the present work should yield  useful 
information for the analysis and prediction of inelastic neutron   
scattering experiments in heavy fermion AF compounds, as CeRh$_2$Si$_2$.

\section{Microscopic model, and perturbative approach.}

\label{model1}


In order to describe the Ce-heavy fermion systems exhibiting
antiferromagnetic long-range order we have used the microscopic model which
has been proposed by Coqblin et al.\cite{coqblin1} to describe the
competition between the Kondo effect and the RKKY interaction, in compounds
where departures from the original Doniach picture\cite{doniach} appear.
In principle, both the RKKY
magnetic coupling and the Kondo effect can be obtained from the Kondo
intrasite-exchange term, but when dealing with approximations it is difficult 
to insure that both effects are taken into account if an explicit intersite 
exchange (as the effective RKKY interaction or, depending on the system, 
also the direct exchange) is not included in the Hamiltonian.\cite{coqblin}    
The model\cite{coqblin1}consists of a Kondo lattice, featuring local
magnetic moments coupled both to conduction electrons, by a Kondo-type
interaction $J_{K}^{{}}$\ , and among themselves by an
antiferromagnetic  RKKY-type exchange $J_{H}^{{}}>0$. The moments are assumed to order
below the N\'eel temperature $T_{N}.$ The Hubbard-correlated conduction
electrons occupy a non-degenerate band. Therefore, the model may be represented
in standard notation by the following Heisenberg-Kondo Hamiltonian: \ 
\begin{eqnarray}
H &=&H_{band}^{{}}+H_{Kondo}+H_{Heis} \\
H_{band}^{{}} &=&\sum_{l\sigma }\epsilon _{l}n_{l\sigma }^{{}}+\
\sum_{l\langle j\rangle \sigma }t_{lj}^{{}}c_{l\sigma }^{\dagger }c_{j\sigma
}^{{}}+U\sum_{l}n_{l\uparrow }^{{}}n_{l\downarrow }^{{}} \\
H_{Kondo} &=&J_{K}^{{}}\left( \sum_{l\in \mathcal{A}}\mathbf{s}%
_{l}^{{}}\cdot \mathbf{S}_{l}^{{}}+\sum_{j\in \mathcal{B}}\mathbf{s}%
_{j}^{{}}\cdot \mathbf{S}_{j}^{{}}\right) \\
H_{Heis} &=&J_{H}^{{}}\sum_{l\langle j\rangle }\mathbf{S}_{l}\mathbf{\cdot S}%
_{j}  \label{sp1}
\end{eqnarray}

Here $l\left\langle j\right\rangle $ $\ $indicates that the lattice site
index $j$\ runs over the $z$\ nearest neighbours of site $l$. The itinerant
electron spin is $s_{l}^{{}}$, while $\mathcal{A}, \mathcal{B}$ 
 label the two interpenetrating
sublattices of local moments, $S_{l\in \mathcal{A}}^{{}}$ and $S_{j\in 
\mathcal{B}}^{{}}$\ , with opposite moment direction. The Kondo exchange $%
J_{K}$\ could in principle have either sign 
(though for heavy fermion compounds, it would be 
antiferromagnetic: $J_{K} \geq 0$).\textit{\ } In the following we shall
distinguish in the Kondo term the longitudinal $H_{K}^{z}$ from the
transverse $H_{K}^{\perp }$ contributions. By taking as positive \ $z$
direction the direction of $S_{l\in \mathcal{A}}^{{}}$\ , they are defined
as: 
\begin{eqnarray}
H_{K}^{z} &=&J_{K}^{{}}\left( \sum_{l\in \mathcal{A}}s_{l}^{z}S_{l}^{z}+%
\sum_{j\in \mathcal{B}}s_{j}^{z}S_{j}^{z}\right)  \\
H_{K}^{\perp } &=&\frac{J_{K}^{{}}}{2}\left[ \sum_{l\in \mathcal{A}}\left(
s_{l}^{+}S_{l}^{-}+H.c.\right) +\sum_{j\in \mathcal{B}}\left(
s_{j}^{+}S_{j}^{-}+H.c.\right) \right]   \notag
\end{eqnarray}

We diagonalize the Heisenberg term, by representing the local moments
operators in the Holstein- Primakoff approximation, which is appropriate at
temperatures much lower than $T_{N}$\ . Namely, we take 
\begin{eqnarray}
\vec{S}_{l} &=&(S_{l}^{+},S_{l}^{-},S_{l}^{z})  \notag \\
&\sim &(\sqrt{2S}b_{l},\sqrt{2S}b_{l}^{\dagger },S-b_{l}^{\dagger }b_{l}) 
\notag \\
\vec{S}_{j} &\sim &(\sqrt{2S}b_{j}^{\dagger },\sqrt{2S}%
b_{j}^{{}},-(S-b_{j}^{\dagger }b_{j})\,,
\end{eqnarray}%
where $b_{l}^{\left( \dagger \right) }$ and $b_{j}^{\left( \dagger \right) }$
are the bosonic operators in sublattices $\mathcal{A}$ and $\mathcal{B}$,
respectively. \ Going to reciprocal space in the reduced Brillouin zone
(RBZ) and introducing the AF spin wave operators $\left\{ a_{q}^{\dagger
},a_{q}^{{}}\right\} $ by the Bogolyubov transformation: 
\begin{eqnarray}
a_{q}^{\dagger } &=&b_{q}^{\dagger }\mathrm{Ch}\left( \vartheta _{q}\right)
-b_{-q}^{{}}\mathrm{Sh}\left( \vartheta _{q}\right)   \notag \\
a_{q}^{{}} &=&b_{q}^{{}}\mathrm{Ch}\left( \vartheta _{q}\right)
-b_{-q}^{\dagger }\mathrm{Sh}\left( \vartheta _{q}\right)  \\
\mathrm{where} &\mathrm{:}&\mathrm{Th}\left( 2\vartheta _{q}\right) =-\frac{1%
}{z}\sum_{\Delta _{lj}}\cos \left( q\Delta _{lj}\right)   \label{bareheis}
\end{eqnarray}%
diagonalizes $H_{Heis}$. In the absence of the interaction with the
fermions, i.e. in the limit of vanishing $J_{K}$, the frequency of the bare AF
spin waves would be ($z$ is the number of nearest neighbors and $\Delta _{lj}
$ is the vector joining two n.n.sites)%
\begin{eqnarray}
H_{Heis} &=&\sum_{q}\hbar \omega _{q}^{{}}\left( a_{q}^{\dagger }a_{q}^{{}}+%
\frac{1}{2}\right)   \notag \\
\hbar \omega _{q}^{{}} &=&zJ_{H}^{{}}S\sqrt{1-\left[ \frac{\sum_{\Delta
_{lj}}\cos \left( q\Delta _{lj}\right) }{z}\right] ^{2}}
\end{eqnarray}

We further assume that the band electrons have developed an AF spin order,
due to the cooperating effects of the Hubbard correlation $U$ and of the
interactions with the AF-ordered local moments, which provide a staggered
field.

To diagonalize $H_{band}$ \ in the AF state, we will use a reformulation of 
Gutzwiller's variational approach for the description of 
 antiferromagnetism in narrow 
 bands due to Spa\l ek et al.\cite%
{SpalekAF} This approach allows to connect the standard (mean-field) Slater
band-insulator to the localized Mott antiferromagnetic insulator.
 One expresses the correlation-induced bandwidth reduction
in the paramagnetic (PM) state by \ a Gutzwiller-type factor $\Phi (n,\eta )$
depending on the band filling $n$ and on the probability of double occupancy 
$\eta =N^{-1}\sum_{l}\left\langle n_{l\uparrow }^{{}}n_{l\downarrow
}^{{}}\right\rangle .$ The correlated band energies $\varepsilon _{k}^{U}$
are written as $\ \varepsilon _{k}^{U}=\Phi (n,\eta )\varepsilon _{k}^{0}$,
where $\varepsilon _{k}^{0}$ are the uncorrelated band energies and:
\begin{equation}
\mathrm{\quad }\Phi (n,\eta )=1-\left( \frac{n}{2-n}\right) \left( 1-\frac{%
4\eta }{n^{2}}\right)   \label{GutzwillerPM}
\end{equation}

The $U$- depending optimal value of $\eta $ is found at zero temperature by
minimizing the PM energy 
\begin{equation*}
E^{PM}=\sum_{k\sigma }\Phi (n,\eta )\varepsilon _{k}^{0}\left\langle
n_{k\sigma }^{{}}\right\rangle +NU\eta 
\end{equation*}%
at given $n$ and $U$ .

Assuming that the electrons have an AF ground state of N\'{e}el-type one 
adopts the standard Slater formalism, only with the PM energies renormalized
according to  Eq.\ref{GutzwillerPM}.

We introduce the fermion operators for this AF Slater-type state $\left\{
\alpha _{k\sigma }^{\left( \dagger \right) },\beta _{k\sigma }^{\left(
\dagger \right) }\right\} $ by the transformation 
\begin{eqnarray}
c_{k\sigma }^{\dagger } &=&\beta _{k\sigma }^{\dagger }\cos \zeta _{k\sigma
}-\alpha _{k\sigma }^{\dagger }\sin \zeta _{k\sigma }  \notag \\
c_{k+\mathbf{\mathcal{Q},}\sigma }^{\dagger } &=&\beta _{k\sigma }^{\dagger
}\sin \zeta _{k\sigma }+\alpha _{k\sigma }^{\dagger }\cos \zeta _{k\sigma }
\end{eqnarray}%
where $\mathbf{\mathcal{Q=}}(1/2,1/2,1/2)$ in units of $2\pi /a$ is the
wavevector characterizing the AF magnetic state and $k$ is a wavevector
belonging to the reduced Brillouin zone (RBZ) defined by $\left\vert \mathbf{%
k}\right\vert \leq \left\vert \mathbf{\mathcal{Q}}/ 2 \right\vert $. The
diagonalization condition for a lattice with an inversion center yields: 
\begin{equation}
\tan \left( 2\zeta _{k\sigma }\right) =-\sigma \frac{U\left\langle
s\right\rangle }{\Phi \varepsilon _{k}}\equiv \sigma \tan \left( 2\zeta
_{k}\right) ,
\end{equation}%
where the AF order parameter $\left\langle s_{l}^{z} \right\rangle \equiv
\left\langle s \right\rangle $  of the itinerant
electrons (staggered magnetization), or band electron polarization, 
 is given by:%
\begin{equation*}
\left\langle s\right\rangle =-\frac{1}{2N}\sum_{k\in RBZ,\sigma }\left(
\left\langle n_{k\sigma }^{\alpha }\right\rangle -\left\langle n_{k\sigma
}^{\beta }\right\rangle \right) \sin 2\zeta _{k}
\end{equation*}%
Therefore, the diagonal bare electron Hamiltonian reads: 
\begin{equation}
H_{band}^{AF}=\sum_{k\in RBZ,\sigma }\left[ E_{k\sigma }^{\alpha }n_{k\sigma
}^{\alpha }+E_{k\sigma }^{\beta }n_{k\sigma }^{\beta }\right] -UN\left( 
\frac{n^{2}}{4}-\left\langle s\right\rangle ^{2}\right) ,  \label{bareband}
\end{equation}%
where, assuming that the lattice has a center of inversion, the bare
electron eigenenergies\ (actually spin-independent) read $\left( x=\alpha
,\beta \right) $: 
\begin{equation}
E_{k\sigma }^{x}=\frac{1}{2}Un+\left( 1-2\delta _{x\alpha }\right) \sqrt{%
\Phi ^{2}\varepsilon _{k}^{2}+\left( U\left\langle s\right\rangle \right)
^{2}},
\end{equation}%
where $\delta _{x\alpha }$ denotes the Kronecker delta. Notice that $\alpha $
is the lower subband. \ The subband filling factors $\left\langle n_{k\sigma
}^{x}\right\rangle =\left\langle n_{k,-\sigma }^{x}\right\rangle =\left[
\exp \left( E_{k\sigma }^{x}-\mu \right) /k_{B}T+1\right] ^{-1}$also depend
on $\left\langle s\right\rangle $ through $E_{k\sigma }^{x}$.

\subsection{Renormalization of the bare subsystems by part of $H_{K}^{z}$}

\label{model3}

By expressing the longitudinal Kondo term in terms of the AF bare electron
operators $\left\{ \alpha _{k\sigma }^{\left( \dagger \right) },\beta
_{k\sigma }^{\left( \dagger \right) }\right\} $ and the bare bosonic
operators $\left\{ a_{q}^{\left( \dagger \right) }\right\} $, it can be
rewritten and decomposed into contributions which are respectively  diagonal in the
bare bosons, i.e. $H_{K}^{z\left( d\right) }$, or non-diagonal, i.e. $I^{z}$.

Explicitly, the diagonal term reads: 
\begin{eqnarray}
H_{K}^{z\left( d\right) } &=&-\frac{J_{K}}{2}S\sum_{k,\sigma }[\sin \left(
2\zeta _{k}\right) \left( n_{k\sigma }^{\alpha }-n_{k\sigma }^{\beta }\right)
\\
&&-\sigma \cos \left( 2\zeta _{k}\right) \left( \alpha _{k\sigma }^{\dagger
}\beta _{k\sigma }^{{}}+\beta _{k\sigma }^{\dagger }\alpha _{k\sigma
}^{{}}\right) ]\left[ 1-V_{q}\right]  \notag
\end{eqnarray}

where the bosonic operator $V_{q}$ reads: 
\begin{eqnarray}
V_{q} &=&\frac{1}{NS}\sum_{q}\left( a_{q}^{\dagger
}a_{q}^{{}}+a_{-q}^{\dagger }a_{-q}^{{}}\right) \mathrm{Ch}\left( 2\vartheta
_{q}\right)  \notag \\
&&+\frac{1}{NS}\sum_{q}\left( a_{q}^{\dagger }a_{-q}^{\dagger
}+a_{q}^{{}}a_{-q}^{{}}\right) \mathrm{Sh}\left( 2\vartheta _{q}\right) 
\notag \\
&&+\frac{2}{NS}\sum_{q}\mathrm{Sh}^{2}\left( \vartheta _{q}\right)
\end{eqnarray}

The term $I^{z}$, non-diagonal in the bare bosons, will be discussed in next
section.

Now, we treat $H_{K}^{z\left( d\right) }$ in mean-field approximation (MFA),
obtaining three kinds of terms. The first one, obtained averaging over the
bare electrons reads: 
\begin{eqnarray}
H_{K}^{z\left( d1\right) } &=&-J_{K}\left\langle s\right\rangle
\sum_{q}\left( a_{q}^{\dagger }a_{q}^{{}}+a_{-q}^{\dagger
}a_{-q}^{{}}\right) \mathrm{Ch}\left( 2\vartheta _{q}\right)  \\
&&-J_{K}\left\langle s\right\rangle \sum_{q}\left( a_{q}^{\dagger
}a_{-q}^{\dagger }+a_{q}^{{}}a_{-q}^{{}}\right) \mathrm{Sh}\left( 2\vartheta
_{q}\right) +\mathrm{const.}  \notag
\end{eqnarray}%
and renormalizes the spin wave frequencies, therefore modifying the
diagonalization condition (Eq.~(\ref{bareheis})) for the AF\ Heisenberg
Hamiltonian as:%
\begin{equation}
\mathrm{Th}\left( 2\vartheta _{q}\right) =-\frac{J_{H}^{{}}S\sum_{\Delta
_{lj}}\cos \left( q\Delta _{lj}\right) }{zJ_{H}^{{}}S-J_{K}^{{}}\left\langle
s\right\rangle }
\end{equation}%
Notice that, in order to minimize the mean-energy of the Kondo term, $%
-J_{K}^{{}}\left\langle s\right\rangle $ is a non-negative quantity,
irrespective of the sign of $J_{K}^{{}}$\ .

The  MFA-redefined\ Heisenberg term now reads: 
\begin{equation}
\bar{H}_{Heis}=\sum_{q}\hbar \Omega _{q}^{{}}\left( a_{q}^{\dagger
}a_{q}^{{}}+\frac{1}{2}\right) 
\label{eq19}
\end{equation}%
with the MFA-renormalized frequency given by: 
\begin{equation}
\hbar \Omega _{q}^{{}}\equiv \left( zJ_{H}^{{}}S+\mid J_{K}^{{}}\left\langle
s \right\rangle \mid \right) 
 \sqrt{1-\left[ \frac{\sum_{\Delta _{lj}}\cos \left(
q\Delta _{lj}\right) }{z+\mid J_{K}^{{}}\left\langle s\right\rangle \mid /J_{H}
^{{}}S}%
\right] ^{2}}
\label{eq20}
\end{equation}%
Averaging over the bare bosons yields two terms. To write them it is
convenient to define the local moment reduction factor as follows: 
\begin{equation}
\mathcal{N}_{sw}\equiv \frac{2}{N}\sum_{q}\left[ \left\langle a_{q}^{\dagger
}a_{q}^{{}}\right\rangle \mathrm{Ch}\left( 2\vartheta _{q}\right) +\mathrm{Sh%
}^{2}\left( \vartheta _{q}\right) \right] 
\end{equation}%
where the mean number of MFA bosons is $\left\langle a_{q}^{\dagger
}a_{q}^{{}}\right\rangle =\left[ \exp \left( \hbar \Omega _{q}/k_{B}T\right)
-1\right] ^{-1}$. \ Then the (in general temperature-dependent) expectation
value of the local moment is $\left\langle S^{z}\right\rangle =S-\mathcal{N}%
_{sw}$.

The second contribution, which reads 
\begin{equation}
H_{K}^{z\left( d2\right) }=-\frac{J_{K}}{2}\left\langle S^{z}\right\rangle
\sum_{k,\sigma }\sin \left( 2\zeta _{k}\right) \left( n_{k\sigma }^{\alpha
}-n_{k\sigma }^{\beta }\right) 
\end{equation}%
renormalizes the electron eigenenergies. The diagonalization condition for
the AF electron Hamiltonian now is:%
\begin{equation}
\tan \left( 2\zeta _{k\sigma }\right) =-\sigma \frac{U\left\langle
s\right\rangle -J_{K}^{{}}\left\langle S^{z}\right\rangle /2}{\Phi
\varepsilon _{k}}
\end{equation}%
Consequently, the AF order parameter $\left\langle s\right\rangle $ of the
itinerant electrons is given by:%
\begin{equation}
\left\langle s\right\rangle =\frac{1}{2N}\sum_{k\sigma }\frac{\left[
U\left\langle s\right\rangle -J_{K}^{{}}\left\langle S^{z}\right\rangle /2%
\right] \left( \left\langle n_{k\sigma }^{\alpha }\right\rangle
-\left\langle n_{k\sigma }^{\beta }\right\rangle \right) }{\sqrt{\Phi
^{2}\varepsilon _{k}^{2}+\left[ U\left\langle s\right\rangle
-J_{K}^{{}}\left\langle S^{z}\right\rangle /2\right] ^{2}}}
\end{equation}%
and the MFA eigenenergies by ($x=\alpha ,\beta $): 
\begin{equation}
E_{k\sigma }^{x}=\frac{1}{2}Un+\left( 1-2\delta _{x\alpha }\right) \sqrt{%
\Phi ^{2}\varepsilon _{k}^{2}+\left[ U\left\langle s\right\rangle
-J_{K}^{{}}\left\langle S^{z}\right\rangle /2\right] ^{2}}
\end{equation}

The third contribution $H_{K}^{z\left( d3\right) }$ provides a hybridization
between the $\alpha ,\beta $ electron states 
\begin{equation}
H_{K}^{z\left( d3\right) }=\frac{J_{K}}{2}\left\langle S^{z}\right\rangle
\sum_{k\sigma }\sigma \cos \left( 2\zeta _{k}\right) \left( \alpha _{k\sigma
}^{\dagger }\beta _{k\sigma }^{{}}+\beta _{k\sigma }^{\dagger }\alpha
_{k\sigma }^{{}}\right) 
\end{equation}%
The transformation 
\begin{eqnarray}
\alpha _{k\sigma }^{\dagger } &=&A_{k\sigma }^{\dagger }\cos \xi _{k\sigma
}+B_{k\sigma }^{\dagger }\sin \xi _{k\sigma }  \notag \\
\beta _{k\sigma }^{\dagger } &=&B_{k\sigma }^{\dagger }\cos \xi _{k\sigma
}-A_{k\sigma }^{\dagger }\sin \xi _{k\sigma }
\end{eqnarray}%
where: 
\begin{equation}
\tan \left( 2\xi _{k\sigma }\right) =\sigma \frac{-J_{K}\left\langle
S^{z}\right\rangle \cos \left( 2\zeta _{k}\right) }{E_{k}^{\alpha
}-E_{k}^{\beta }-J_{K}\left\langle S^{z}\right\rangle \sin \left( 2\zeta
_{k}\right) }
\end{equation}%
allows to eliminate $H_{K}^{z\left( d3\right) }$ yielding the
MFA hybridized electron Hamiltonian in diagonal form as: 
\begin{equation}
\bar{H}_{band}^{AF}=\sum_{k\sigma }\left[ \mathcal{E}_{k\sigma
}^{B}n_{k\sigma }^{B}+\mathcal{E}_{k\sigma }^{A}n_{k\sigma }^{A}\right] 
\end{equation}%
The hybridized band energies are ($X=A,B$) 
\begin{equation}
\mathcal{E}_{k}^{X}=\frac{1}{2}\left( E_{k}^{\alpha }+E_{k}^{\beta }\right)
-\left( 2\delta _{XA}-1\right) \frac{\Gamma _{k}}{2}
\label{hybbands}
\end{equation}%
where%
\begin{eqnarray}
\Gamma _{k} &=&\left\vert E_{k}^{\alpha }-E_{k}^{\beta }\right\vert \times  
\notag \\
&&\times \sqrt{\left[ 1-\frac{J_{K}\left\langle S^{z}\right\rangle \sin
\left( 2\zeta _{k}\right) }{E_{k}^{\alpha }-E_{k}^{\beta }}\right] ^{2}+%
\left[ \frac{J_{K}\left\langle S^{z}\right\rangle \cos \left( 2\zeta
_{k}\right) }{E_{k}^{\alpha }-E_{k}^{\beta }}\right] ^{2}}  \notag \\
&&
\label{hyb2}
\end{eqnarray}%
is the difference of energy at wavector $k$ between the AF subbands. 
Notice that the $A$ band is the lower one. Taking the
paramagnetic band filling per site $n\leq 1$ only this band will be
non-empty.

\subsection{Perturbative derivation of the effective magnon Hamiltonian.}

\label{model3b}

At this stage, we will describe the perturbative treatment performed to
derive the effective second-order Hamiltonian for magnons. We begin by
rearranging the Hamiltonian terms, splitting it into a basic part $H_{0}$,
including the MFA redefined \textquotedblleft bare\textquotedblright\ magnon
and electron Hamiltonians of previous section, plus a \textquotedblleft
perturbation\textquotedblright\ $I$, which includes the full transverse
Kondo coupling term ($I^{\perp }\equiv H_{K}^{\perp }$) and the scattering
(i.e. non-diagonal in boson operators) part of the longitudinal Kondo
coupling ($I^{z}$) not considered so far.

Explicitly, we have $H=H_{0}+I=H_{0}+I^{z}+I^{\perp }$ where \ $H_{0}=\bar{H}%
_{Heis}+\bar{H}_{band}^{AF}$. The longitudinal perturbation $I^{z}$ reads: 
\begin{eqnarray}
I^{z} &=&\frac{J_{K}}{N}\sum_{pqr,\sigma }\left( 1-\delta _{pq}\right)
M_{r,p+r-q}^{++}\times   \notag \\
&&\times \left( A_{r\sigma }^{\dagger }A_{p+r-q,\sigma }^{{}}-B_{r\sigma
}^{\dagger }B_{p+r-q,\sigma }^{{}}\right) G_{pq}  \notag \\
&&-\frac{J_{K}}{N}\sum_{pqr,\sigma }\sigma \left( 1-\delta _{pq}\right)
L_{r,p+r-q}^{++}\times   \notag \\
&&\times \left( A_{r\sigma }^{\dagger }B_{p+r-q,\sigma }^{{}}+B_{r\sigma
}^{\dagger }A_{p+r-q,\sigma }^{{}}\right) G_{pq}  \label{longitudinal_pert}
\end{eqnarray}%
The spin wave operators are contained in 
\begin{eqnarray*}
G_{pq} &=&a_{p}^{\dagger }a_{q}^{{}}\mathrm{Ch}\left( \vartheta _{q}\right) 
\mathrm{Ch}\left( \vartheta _{p}\right) +a_{-q}^{\dagger }a_{-p}^{{}}\mathrm{%
Sh}\left( \vartheta _{q}\right) \mathrm{Sh}\left( \vartheta _{p}\right)  \\
&&+a_{p}^{\dagger }a_{-q}^{\dagger }\mathrm{Sh}\left( \vartheta _{q}\right) 
\mathrm{Ch}\left( \vartheta _{p}\right) +a_{-p}^{{}}a_{q}^{{}}\mathrm{Ch}%
\left( \vartheta _{q}\right) \mathrm{Sh}\left( \vartheta _{p}\right) 
\end{eqnarray*}%
while coefficients $M_{r,p+r-q}^{++}$ and $L_{r,p+r-q}^{++}$ are defined
in the Appendix.

The transverse perturbation is written as:\ 
\begin{eqnarray}
I^{\perp } &=&J_{K}\sqrt{\frac{S}{2}}\sum_{q}\left\{ \left[ a_{q}^{\dagger }%
\mathrm{Ch}\left( \vartheta _{q}\right) +a_{-q}^{{}}\mathrm{Sh}\left(
\vartheta _{q}\right) \right] \left( s_{q_{A}}^{+}+s_{qB}^{-}\right) \right. 
\notag \\
&&\left. +\left[ a_{q}^{\dagger }\mathrm{Sh}\left( \vartheta _{q}\right)
+a_{-q}^{{}}\mathrm{Ch}\left( \vartheta _{q}\right) \right] \left(
s_{-qB}^{+}+s_{-q_{A}}^{-}\right) \right\}   \label{transverse_pert}
\end{eqnarray}

The spin operators are expressed through the electron operators as: 
\begin{eqnarray}
s_{q_{A}}^{\lambda }+s_{q_{B}}^{\tau } &=&  \notag \\
&=&\frac{1}{2}\sqrt{\frac{2}{N}}\sum_{k\sigma }A_{k\sigma }^{\dagger
}A_{k+q,-\sigma }^{{}}\mathcal{C}_{AA}^{\lambda \tau }\left( k,q\right)  
\notag \\
&&+\frac{1}{2}\sqrt{\frac{2}{N}}\sum_{k\sigma }B_{k\sigma }^{\dagger
}B_{k+q,-\sigma }^{{}}\mathcal{C}_{BB}^{\lambda \tau }\left( k,q\right)  
\notag \\
&&+\frac{1}{2}\sqrt{\frac{2}{N}}\sum_{k\sigma }\sigma A_{k\sigma }^{\dagger
}B_{k+q,-\sigma }^{{}}\mathcal{C}_{AB}^{\lambda \tau }\left( k,q\right)  
\notag \\
&&+\frac{1}{2}\sqrt{\frac{2}{N}}\sum_{k\sigma }\sigma B_{k\sigma }^{\dagger
}A_{k+q,-\sigma }^{{}}\mathcal{C}_{BA}^{\lambda \tau }\left( k,q\right)  
\notag \\
&&  \label{sA+sB}
\end{eqnarray}%
where $\lambda ,\tau =\pm $ and the coefficients $\mathcal{C}_{XY}^{\lambda
\tau }\left( k,q\right) $ with $X,Y=A,B$ are defined in the Appendix.




In the previous section, we have included at  MFA level the effects of the diagonal part
of the longitudinal Kondo coupling on the bosonic states
(Eqs.~\ref{eq19}-\ref{eq20} )  
and on the AF electronic states (Eqs.~\ref{hybbands}-\ref{hyb2} ). 
We will now determine the magnon renormalization effects derived from
the transverse and non-diagonal parts of the longitudinal Kondo coupling,
which we are considering as the perturbation $I$ on the MFA state.

The effect of the perturbation will be taken into account through a Fr\"{o}%
hlich-type of truncated unitary transformation \cite{wagner}. We determine
the generator $R$ of an appropriate canonical transformation by eliminating
the first order term in the perturbation. To this aim, we impose $%
I+i[R,H_{0}]=0$. Introducing the notation $\mathcal{R\equiv R}^{z}+\mathcal{R}^{\perp }$,
we decompose this constraint into two separate equations, which can be
solved yielding: $\mathcal{R}^{z\left( \perp \right) }=\lim_{t\rightarrow 0}%
\frac{i}{\hbar }\int_{-\infty }^{t}I^{z\left( \perp \right) }(x)dx.$ By this
procedure,\cite{tlpaper} we obtain the second order effective Hamiltonian for the
magnon-conduction electron system as: 
\begin{eqnarray}
H_{eff} &=&\bar{H}_{band}^{AF}+\bar{H}_{Heis}+\frac{1}{2}\left[ \mathcal{R}%
^{z}+\mathcal{R}^{\perp },I^{z}+I^{\perp }\right]  \notag \\
&&+\mathcal{O}\left( I^{z}+I^{\perp }\right) ^{3}\quad , \label{controlparameter}
\end{eqnarray}%
Let us remark here that the perturbative parameter which actually controls
this expansion, is the ratio $J_{K}/J_{H}$ weighed by factors(coefficients)
which depend on the electronic band structure and filling, as can be seen
from the explicit expressions for $I^{z}$ and $I^{\perp }$ given above.
These electronic coefficients effectively reduce the magnitude of the
perturbative control parameter from the raw value $J_{K}/J_{H}$, leading to
a smooth convergence of the perturbative expansion even when $\mid
J_{K}/J_{H}\mid $
is near or exceeds one. This will become clear when we present our results
for the renormalized magnons in next section.

Each term in the perturbation produces a corresponding term in the
generator. From $I^{z}$\ we obtain the "longitudinal" generator $R^{z}$\
while from $I^{\perp }$\ we obtain the \bigskip "transverse" generator $%
R^{\perp }$\ , which are conveniently decomposed as:%
\begin{equation}
R^{z}=\sum_{X,Y=A,B}\sum_{m=1,4}R_{mXY}^{z}\quad R^{\perp
}=\sum_{X,Y=A,B}R_{XY}^{\perp }  \label{R_decomposition}
\end{equation}

The terms $R_{mXY}^{z}$\ and $R_{XY}^{\perp }$\ \ are detailed in the
Appendix.

Finally, we make a projection onto the AF fermion wavefunction to obtain a
second-order effective Hamiltonian for the magnons $H_{SW}^{eff}$. When
taking the average $\left\langle H_{eff}\right\rangle _{Fermi}$over the AF\
Fermi wavefunction we find 
\begin{equation}
\left\langle \left[ \mathcal{R}^{z},I^{\perp }\right] \right\rangle
_{Fermi}=\left\langle \left[ \mathcal{R}^{\perp },I^{z}\right] \right\rangle
_{Fermi}=0
\end{equation}%
so that the effective spin wave Hamiltonian has the simpler form%
\begin{eqnarray}
H_{SW}^{eff} &\equiv &\left\langle H_{eff}\right\rangle _{Fermi}=  \notag \\
&=&\sum_{k\sigma }\left( \mathcal{E}_{k\sigma }^{A}\left\langle n_{k\sigma
}^{A}\right\rangle +\mathcal{E}_{k\sigma }^{B}\left\langle n_{k\sigma
}^{B}\right\rangle \right) +\sum_{q}\hbar \Omega _{q}a_{q}^{\dagger
}a_{q}^{{}}  \notag \\
&&+\frac{1}{2}\left\langle \left[ \mathcal{R}^{z},I^{z}\right] +\left[ 
\mathcal{R}^{\perp },I^{\perp }\right] \right\rangle _{Fermi}
\label{H_eff_SW_a}
\end{eqnarray}

where \ 
\begin{equation}
\left\langle n_{k\sigma }^{X}\right\rangle =\left[ \exp \left( \mathcal{E}%
_{k\sigma }^{X}-\mu \right) /k_{B}T+1\right] ^{-1}
\end{equation}

Evaluating \ $\frac{1}{2}\left\langle \left[ \mathcal{R}^{z},I^{z}\right] +%
\left[ \mathcal{R}^{\perp },I^{\perp }\right] \right\rangle _{Fermi}$\ we
arrive at: 
\begin{eqnarray}
H_{SW}^{eff} &=&\sum_{q}\hbar \left( \Omega _{q}+\Theta _{q}\right)
a_{q}^{\dagger }a_{q}^{{}}+\sum_{q}\hbar \Psi _{q}\left( a_{q}^{\dagger
}a_{-q}^{\dagger }+a_{q}^{{}}a_{-q}^{{}}\right)  \notag \\
&&+\mathrm{const.}  \label{H_eff_SW_b}
\end{eqnarray}

The coefficients $\hbar \Theta _{q}$\ and $\hbar \Psi _{q}$\ are given by
long and complicated expressions, which we detail in the Appendix.\ Here we
just point out that both the harmonic and the anharmonic parts in $%
H_{SW}^{eff}$\ have contributions from both longitudinal and transverse
Kondo terms.

With one last Bogolyubov transformation, we diagonalize the effective
Hamiltonian for the spin excitations, yielding: \ 
\begin{eqnarray}
H_{SW}^{eff} &=&\sum_{q}\hbar \left[ \left( \Omega _{q}+\Theta _{q}\right) 
\sqrt{1-\frac{4\Psi _{q}^{2}}{\left( \Omega _{q}+\Theta _{q}\right) ^{2}}}%
\right] a_{q}^{\dagger }a_{q}^{{}}  \notag \\
&\equiv &\sum_{q}\hbar \widetilde{\Omega }_{q}a_{q}^{\dagger }a_{q}^{{}} \label{H_eff_final}
\end{eqnarray}
where $\widetilde{\Omega }_{q}$ is the renormalized antiferromagnetic spin
wave frequency.


\section{Results and discussion}

\label{results}

In the previous section, we obtained a  formally simple final expression 
for the renormalized antiferromagnetic magnons in 
Eq.~(\ref{H_eff_final}),  but which depends on a series
of coefficients which are  detailed in the Appendix. 
The long complicated expressions for these perturbatively obtained coefficients,
depend on the combined effect or interplay  of the different model parameters: 
 in particular, some coefficients involve double and triple summmations,  
over the reduced Brillouin zone, of q-dependent filling factors of the 
conduction electron bands. Therefore, we have evaluated our renormalized 
magnon results numerically,   exploring  wide ranges of the 
different model parameters.  This has allowed us not only to   
assess and compare the influence and  main effect of each of the different 
model parameters, and show that one can reasonably explain 
experimental magnon results \cite{cepd2si2,cein3} employing parameters 
in the range independently shown to be appropriate for a phenomenological fit 
of specific heat measurements in this family of antiferromagnetic 
heavy fermions,\cite{lobos}  as we will show below. 
Our exploration of wide parameter ranges 
 has also allowed us to verify that the convergence radius 
of the perturbative series which determines the magnon renormalization 
is much wider than one might naively have expected. As mentioned 
below Eq.~(\ref{controlparameter}), results to be shown in this section    
evidentiate that the actual ``small parameter'' controlling this perturbative 
expansion is not just the bare $J_{K}/J_{H}$ ratio: this appears weighed 
by electronic structure and filling dependent coefficients 
which effectively  reduce the  control parameter value  from this ratio, 
leading to convergent magnon renormalization results for $\mid J_{K}/J_{H}
\mid > 1 $,  
when combined with suitable values of the other model parameters. 

The numerical study of our model has been done assuming
two simple cubic interpenetrating magnetic sublattices, for simplicity
and without much loss of generality  (rigorously, under this
assumption our results would correspond to a cubic (bcc) lattice). 
Notice  that CeIn$_3$,\cite{cein3} one of the compounds where inelastic neutron scattering (INS) on single crystals has measured the 
magnons we aim to describe, is cubic (though fcc) 
and has a three-dimensional N\'eel-type antiferromagnetic structure.  
We evaluated  magnon excitations at zero temperature,  and 
assuming  an underlying 3D N\'eel-type antiferromagnetic ground
state of the system: our study focuses on parameter sets far away from the 
quantum critical region of these systems,  
deep inside the antiferromagnetic phase.  
In fact, our parameters lie well inside the AF stable region recently 
determined  by a  DMFT + NRG study\cite{pruschke}  
 of the magnetic phase diagram  of the correlated Kondo-lattice 
(corresponding to the $J_{H}=0$ case of our model:  the addition of 
non-negligible AF-like RKKY coupling $J_H$, as done here,  
will only increase the stability of the AF phase). 
As expt. observed,\cite{cein3} in this range one might expect  Kondo-type 
spin fluctuations  to be less relevant and  the dispersive spin waves,  
object of our study, to appear in the AF phase.  
In accordance with expt. indications\cite{cein3}  and for simplicity, we will
further assume there is one isolated crystal field level of Ce$^{3+}$  
which is relevant, hosting a spin $ S = 1/2 $ (in fact, $m = 0.5 \mu_{B}$ is
the exp. magnitude of the local moments in CeIn$_3$\cite{cein3} ) 
, and concentrate on the
numerical evaluation of the renormalized spin waves given by  Eq.~(\ref{H_eff_final}). 
Hopping parameter $t$ is taken as unit of energy, being $W=12 t$ the
total bare electron bandwidth.  

In the following, we will start by exhibiting and discussing the general
trends we have found in our study of magnon renormalization in the 
Heisenberg-Kondo model with a correlated conduction band. 
We shall end this section by focusing on the description of the measured 
magnons in antiferromagnetic heavy fermion compounds.

\begin{figure}[t]
\begin{picture}(0,0)%
\includegraphics{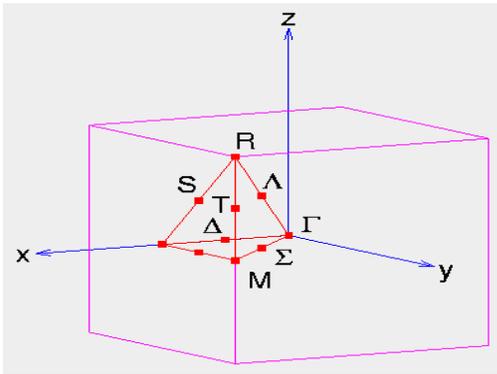}%
\end{picture}%
\setlength{\unitlength}{1243sp}%
\begingroup\makeatletter\ifx\SetFigFont\undefined%
\gdef\SetFigFont#1#2#3#4#5{%
  \reset@font\fontsize{#1}{#2pt}%
  \fontfamily{#3}\fontseries{#4}\fontshape{#5}%
  \selectfont}%
\fi\endgroup%
\begin{picture}(9900,7380)(811,-7261)
\end{picture}%
\caption{Simple cubic lattice: 1st Brillouin zone and special symmetry
  points.}
\label{crystal}
\end{figure}

Fig.~(\ref{crystal}) depicts the first Brillouin zone (BZ) of the 
 simple cubic lattice,
and we include the notation for the special symmetry points and BZ paths 
on which the spin waves were numerically evaluated 
($\Gamma \equiv O$ denotes the zone center, 
$\Delta \equiv X= 0.5 \pi /a \, (1,0,0)$, $Y= 0.5 \pi /a \,
 (0,1,0)$),$Z= 0.5 \pi /a \, (0,0,1)$ ). 
Let us  mention here that for the BZ summations we have used the special
points BZ sampling method by Chadi-Cohen (CC), at 4th order,  
for the simple cubic lattice.\cite{macot} To obtain dressed magnons 
with the correct symmetry of the lattice, we have noticed \cite{tlpaper}
 it is not sufficient  to use 
the basic set of wave vectors of the (reduced) 1st BZ octant, but one
needs to extend it to the (reduced) full 1st BZ (to properly take into
account points at the frontiers between octants): we do this  
applying the (48) symmetry operations of the $O_h$ group 
to the basic set.
Thus, at 4th-order CC, taking into account all symmetry operations we
 have  included 5760 special symmetry points for each BZ summation. 
To achieve higher accuracy for our determination of the Fermi level 
(a delicate issue close to half-filling), we have used integration over 
the 5th order Chadi-Cohen vectors ( that is, we used 39168 special 
symmetry points taking all symmetry operations into account).

\begin{figure}[h]
\includegraphics[angle=270 , width=\columnwidth]{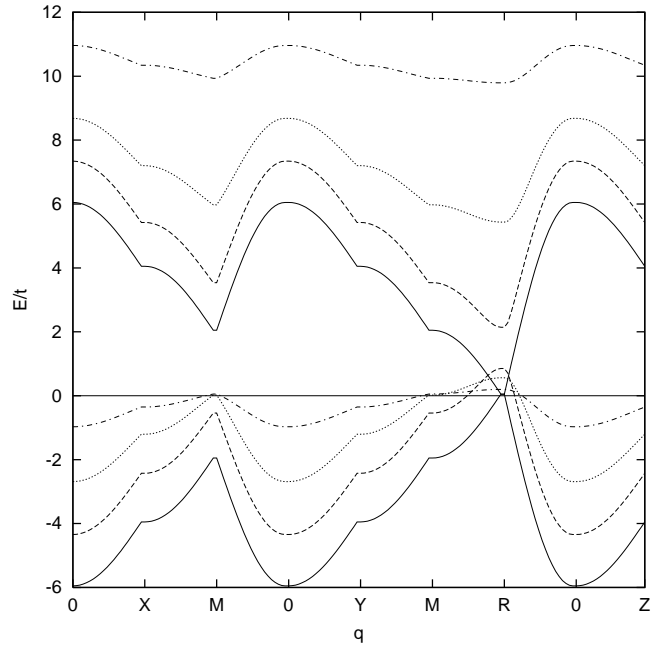} 
\caption{Hybridized electron bands ($\uparrow$) along selected BZ
paths.  Parameters: $S=1/2; T=0; t=1eV ; J_H/t=0.001; J_K/J_H =1;
n=0.999 $.
$ U/t=0.1 $ (full lines), 3 (dashed), 6 (dotted), 10 (dash-dotted).}
\label{bands}
\end{figure}

In Fig.~(\ref{bands})  we show the typical conduction electron
bandstructure near half-filling,   given by Eq.~(\ref{hybbands}), 
obtained with the variational approach described in section~(\ref{model3}). 
Notice that for each value of $U$ only the lower band (denoted $A$) is filled.
Chemical potential values obtained for the four cases shown are: 
$\mu/t = -0. 63, 0.58, 0.49, 0.18$, (slightly below the top of the
lowest band), respectively for $U/W= 0.008, 0.25, 0.5, 0.83$ 
[ i.e.  $U/t$ values of Fig.~(\ref{bands}) ]. We obtain a direct band
gap, determined by the AF subband energies at the cube diagonal BZ
point: $q=R$ (see Fig.~(\ref{crystal}) ). As expected, the size of the 
gap, as well as the energy of its centroid,  increase  with 
 electron correlation ($U$) magnitude, while also a     
correlation-driven band narrowing effect is being described.  
The AF band gap value essentially 
depends on $U\left\langle s \right\rangle $,  
with $\left\langle s \right\rangle $ growing with both the band
filling $n$ and correlation $U$. In Table~\ref{table1} these trends 
of the AF band polarization with the different parameters are recognized. 
It is also interesting to compare  our AF band polarization values
(tabulated) with those reported in Fig.2 of Ref.~[\cite{pruschke}], 
for corresponding parameters. 
Notwithstanding the different respective treatments  for the band electron
correlations, we find quite reasonable agreement 
where we could check it: in the present work we have used 
antiferromagnetic Kondo coupling values in a relatively narrow region 
around AF $J_K/W \sim 10^{-5}-10^{-3}$. 
In agreement with Peters and Pruschke,\cite{pruschke} 
in this parameter range we find that 
the local moments are almost fully polarized,  
while the corresponding band polarization obtained for $U \sim 0$ and $U \sim W$ 
agrees quite well with the corresponding values in Fig.2 of Ref.~[\cite{pruschke}]. 

\begin{figure}[h]
\includegraphics[angle=270 , width=\columnwidth]{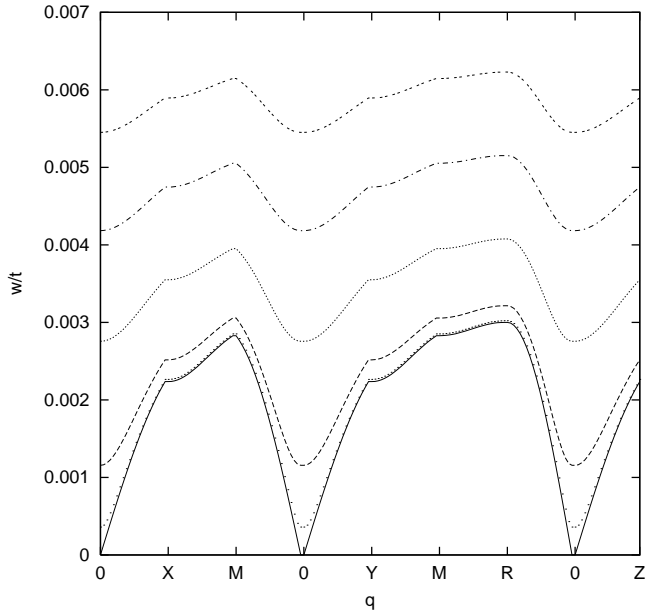} 
\caption{$J_k/J_H$ dependence of renormalized AF magnons: magnon energy along
  selected BZ paths. Parameters: $ S=1/2; T=0; t=1eV ; J_H/t=0.001 ;
n=0.999 ; U/t = 3$. Full line: bare AF magnons. 
$J_K/J_H$ = 0.1 (sparsely dotted line), 1 (dashed), 5 (densely dotted), 
10 (dash-dotted), 15 (double-dashed). }
\label{jratiodep}
\end{figure}

In Fig.~(\ref{jratiodep}) we show the dependence of the AF magnon renormalization 
on the bare $J_K/J_H$ ratio at half-filling ( the most relevant filling 
for AF heavy fermion compounds):  the lowest curve represents 
the bare magnons $\omega_{q}$ (independent of the conduction electrons).  
As a general trend, we find that the renormalization effects beyond mean-field
approximation  reduce the spin wave frequency $ \widetilde{\Omega}_{q}$ 
with respect to the MFA value $\Omega_{q}$, 
but still the total renormalization effect 
$ \omega_{q}\Longrightarrow\widetilde {\Omega}_{q}$ 
is a hardening with respect to the bare (non-interacting) magnon
frequency: $\omega_{q}<\widetilde{\Omega}_{q}<\Omega_{q}$. 
It is reasonable that due to their coupling to the conduction electrons, 
magnon excitations  in the coupled magnetic system  require a larger
energy than 
when the excitation   involves only  the local moment subsystem. Thus,   
 the dressed magnon energies increase with respect to the bare ones in an
amount  proportional  
to the increase in relative weight of the Kondo coupling 
with respect to the bare RKKY (Heisenberg) coupling, as  Fig.~(\ref{jratiodep}) shows.
Meanwhile, $ z J_H S $  sets the main energy scale for the magnons. On
the other hand, the renormalization due to only mean-field approximation  
effects is a hardening with respect to bare magnons:  notice that the
MFA of the longitudinal part of the Kondo interaction employed here, leads to 
a factor of $z J_H S + \mid J_K <s> \mid $  for the ``MFA-renormalized'' frequency,
from which the (longitudinal) Kondo coupling is seen to induce a magnon
hardening: which is partially reduced when the  full renormalization effects   
are taken into account.

\begin{figure}[h]
\includegraphics[angle=270 , width=\columnwidth]{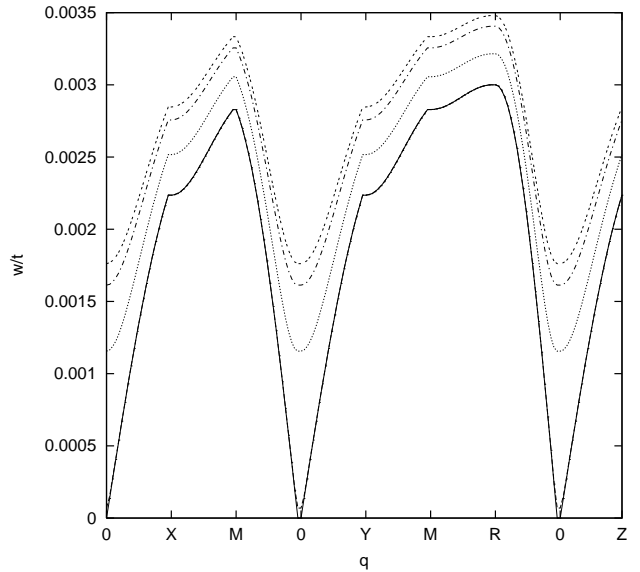} 
\caption{$U$ dependence of renormalized AF magnons: magnon energy along
  selected BZ paths. Parameters: $U$ as detailed here;
others as in Fig. 2.  Full line: bare AF magnons. 
$U/t = 0.1$ (sparse dots: indiscernible from bare magnon curve in
plot), 1 (dashed line), 3 (densely dotted), 6 (dash-dotted) , 10 (double-dashed).}
\label{udep}
\end{figure}

Next figure ( Fig.~(\ref{udep}) ) depicts the dependence of the renormalized 
magnons on the correlations in the conduction band ($U$),  at half-filling: 
the trend is a hardening of magnons when $U$ is increased 
( such hardening appears also for the MFA
magnons), and we show cases where $U/W$ ranges between 0.008 (for $U/t =0.1$) and
0.83 ($U/t = 10$). This may be understood by the reinforcement of 
U-dependent antiferromagnetism (spin polarization) in the conduction band,  
 which renders 
conduction electrons less available  to follow easily the spin excitation 
determined by magnon excitation in the local moment 
subsystem: an energetic cost is involved. In fact, 
the magnon gap value we obtain at $q=0$ is directly related to the correlation value, 
and to the symmetry breaking involved in our treatment when 
the itinerant antiferromagnetism of the conduction band is considered 
( as was also reported in previous work on spin excitations in manganites).
\cite{golosovU} Notice that increasing $U$ 
also reduces q-dependent details in the renormalized magnons: 
this can be explained  by analizing the indirect effect 
which  $U$  has on magnons  (while $J_K$ has also  a direct effect, 
since it appears  also as explicit multiplicative factor 
of the perturbatively obtained magnon corrections). 
 $U$ affects magnons through the modifications it induces  
in electron bandstructure, and in particular the $U$-driven increase 
of energy denominators  of the perturbative coefficients which
determine  the q-dependent renormalization of magnons (details in the
Appendix), which explains the above mentioned result. 
 The recent more refined DMFT+NRG treatment of
correlations in an extended  Kondo lattice model\cite{pruschke}  unfortunately 
does not allow us comparison, here, as their  
finite U results are presented for antiferromagnetic  $J_K$  outside the region 
of interest in our problem:  their correlated AF Kondo coupling system  
is studied at much too large Kondo coupling (namely, $J_K = 0.5 W =
U$) for the antiferromagnetic state to remain stable, 
 being  the Kondo insulator with all moments locally quenched 
the stable phase near half filling in that case. 
 
\begin{figure}[h]
\includegraphics[angle=270 , width=\columnwidth]{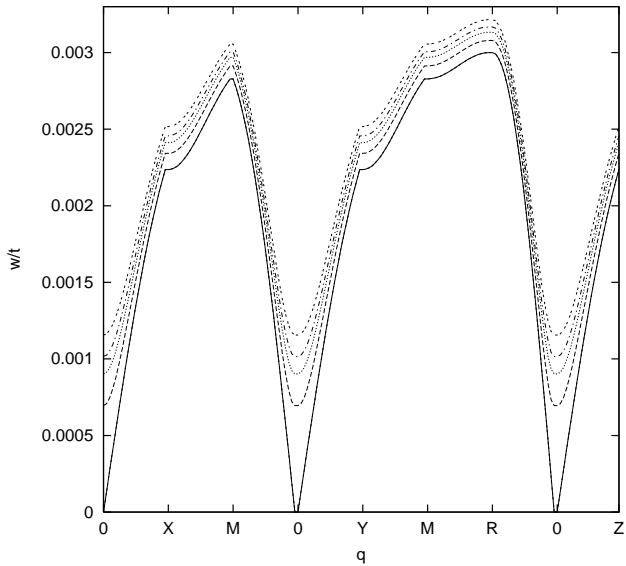} 
\caption{Filling $(n)$ dependence of renormalized AF magnons: magnon
energy along selected BZ paths. Parameters: $U/t= 3$; $n$ as detailed here;
others as in Fig. 2. Full line: bare AF magnons. $n = 0.999$ 
(double-dashed line), 0.8 (dash-dotted), 
0.7 (dotted), 0.6 (dashed), 0.4 (sparsely dotted line: indiscernible
from bare magnon curve in plot).}
\label{ndep}
\end{figure}

We exhibit effects of the doping on the magnon renormalization 
in Fig.~(\ref{ndep}).  Here the deviation of the renormalized magnon energies 
from the bare magnon values is increased with the filling: at half-filling 
the renormalization is largest, there being  more
conduction electrons present, which contribute to the renormalization of magnons 
by their Kondo coupling to the local moments. Doping away from half-filling 
we  obtain a smooth reduction of such renormalization effects. Thus, both
filling and electron correlation do increase renormalization effects, 
and we have already mentioned that both result 
in similar increases of spin polarization 
of the conduction band. We will shortly come back to this point, when
introducing our last figure.

Let us briefly refer again to the q-dependence of the AF magnon
renormalization we find. Some anisotropy is present: a larger q-dependence 
is noticeable along  BZ diagonal paths such as $O-M$ or $O-R$ (see e.g. 
 Fig.~(\ref{ndep})) or paths along the symmetry axes. While the
 renormalization effects are more pronounced at long wavelenghts: in
 particular, they are maximal at the BZ center 
where we find a spin stiffness increasing with doping,  and decreasing with U
or $J_K/J_H$.  Making allowance for the quite different systems involved, 
let us mention that the renormalized AF magnon behavior 
we obtain 
contrasts with the one recently disclosed  by INS measurements in {\it ferromagnetic 
metallic manganites}:\cite{ye} where low-q   spin wave stiffness appears 
insensitive to doping, 
while magnons exhibit a doping-dependent renormalization at the BZ boundaries
(recently suggested to be related to electronic correlations\cite{kapetanakis}).

\begin{figure}[h]
\includegraphics[angle=270 , width=\columnwidth]{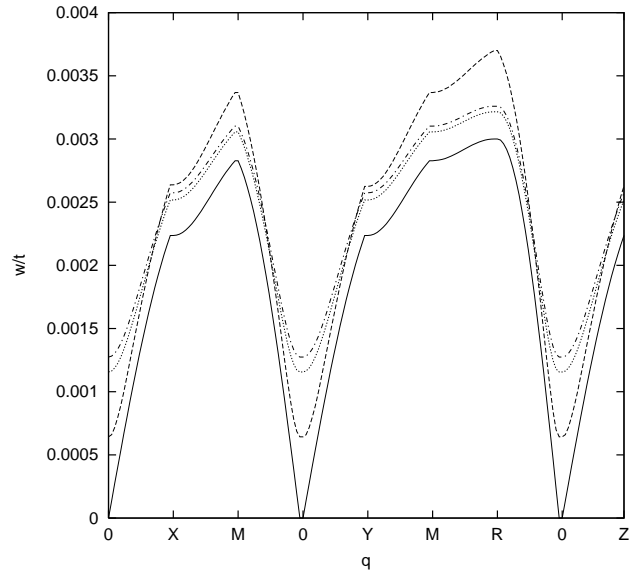} 
\caption{ Three parameter sets leading to renormalized AF magnons as
in CeIn$_3$. (a): $J_K/J_H =100 $, U/t = 0.001, n= 0.999 ( bare magnons: full
line; ren. magnons: dashed); (b): $J_K/J_H= 1$, U/t= 3, n= 0.999 (dotted); 
(c): $J_K/J_H= 1$, U/t= 6, n= 0.7 (dash-dotted). Other parameters as Fig. 2. }
\label{nearcein3}
\end{figure}

 At this point, let us compare our results with the few INS magnon
measurements available for single crystals 
of antiferromagnetic heavy fermions. Comparison in more detail 
may be made only with CeIn$_3$, which is  
 cubic (though f.c.c.) and presents a three-dimensional AF order 
as we have assumed for our calculation. 
In Fig.~(\ref{nearcein3}) we show magnon results we obtained with three
 different sets of model parameters: as can be seen,   
all of them providing a reasonably good description 
of the measured CeIn$_3$ magnons.\cite{cein3} At the same time, 
these parameter sets yield examples of behavior of the 
magnon renormalization of the model discussed above: 

(a) the first parameter  set used in Fig. 6 provides an example of convergence of our
 perturbative expansion for a case with $J_K > J_H (=100 J_H)$, at
 half-filling and with negligible $U$; notice a slightly larger q-dependence 
is obtained in this case (to be expected from the q-dependent
 expressions of the perturbative coefficients): e.g.  easily  noticeable 
  different height of the main peaks (at symmetry points M and R) is visible; 

(b) with parameter set (b) of Fig. 6, again a 
reasonable qualitative description  of expt. CeIn$_3$ magnons is obtained 
 (considering our calculations
are done for sc magnetic sublattices (thus bcc lattice), 
while CeIn$_3$ in fact is fcc). The agreement here is slightly better 
than with set (a). This shows how delicate 
the interplay between the different parameters of the model is:  
since, compared with parameter set (a),  here we obtain 
a quite similar result at half-filling  but 
with   $J_K= J_H $ and $ U/t = 3$; 

(c) finally 
we illustrate that we can obtain a quite similar description of the 
CeIn$_3$ measured magnons away from half filling: e.g. assuming $n=0.7$ but  
compensating the decrease in electron energy due to doping 
 by an increase in band correlation: 
taking $U/t = 6$  while keeping $ J_K = J_H $.

It is also worth mentioning that these sets of parameters allowing description
of INS results are in parameter ranges independently suggested 
by other authors for these compounds.  
A concrete example  is the   phenomenological fit of specific heat curves 
for CeIn$_3$ , which was made by Lobos et al.,\cite{lobos} using:
 $J_H/t = 0.0014$, $t=0.5eV$, $n= 1$  and $J_K/J_H = 980$; 
 while for CeRh$_2$Si$_2$ they have used:  $J_H/t = 0.0034$ and $J_K/J_H = 430$; 
and their data extrapolation  for  CePd$_2$Si$_2$ was: 
$J_H/t = 0.0034$ with a negligible $J_K/J_H$.

\section{Summary}

\label{conclusions}

In the present work, we have studied spin wave excitations  in heavy fermion 
compounds  with antiferromagnetic long-range order , 
  where a  strong competition of RKKY and Kondo screening is present, 
as evidenced by nearly equal magnetic ordering and Kondo temperatures. We
have described these systems using a microscopic model including a lattice of 
correlated f-electron orbitals (as in Ce-, U- compounds of this family) 
hybridized with a correlated conduction band, with the presence of competing 
RKKY-Heisenberg and Kondo magnetic couplings.

Through a series of unitary transformations we perturbatively derived a 
second-order effective Hamiltonian describing the spin wave excitations,
renormalized by their interaction with the conduction electrons. We have 
numerically studied the effect of the different parameters of this effective
model on the magnon energy renormalization by the conduction electrons.

We have been able to find appropriate sets of model parameters to describe 
the few existing measurements of magnons by inelastic neutron scattering in
single crystal samples of antiferromagnetic heavy fermion Ce compounds. These 
parameter sets also agree with the ranges proposed for these compounds, 
through phenomenological fits of other experiments, like specific heat.
Our results may provide information of interest for the prediction of 
inelastic neutron scattering experiments in other compounds of this family,
like CeRh$_2$Si$_2$, where there have been suggestions that the  RKKY coupling 
should be stronger than Kondo effect,\cite{severing} 
and  the only existing NIS measurements are of poor quality: 
they were made on polycrystals\cite{severing} many years ago, 
and with lower resolution. 
    
As outlook towards related future work, we might mention the description 
of the experimentally reported magnon damping effects,  
and the study of the  coexistence of antiferromagnetism and superconductivity 
in the context of the present model.


\acknowledgments
We thank J. Sereni, E. M\"uller-Hartmann, G. Aeppli, P. Coleman, P.
Gegenwart, P. Santini, A. Lobos, and J.R. Iglesias for discussions and
references. M.A. thanks Centro At\'omico Bariloche for the hospitality 
and support. C.I.V. is Investigador Cient\'{\i}fico of CONICET(Argentina), 
and acknowledges support from CONICET (PEI'6298 and PIP'5342 grants), 
Consiglio Nazionale delle Ricerche (CNR Short-Term Mobility Grant 2006), 
as well as hospitality and support from Dipto. di Fisica (Univ. di Parma), 
Inst. f\"ur Theoretische Physik (Univ. zu K\"oln) 
and the International Centre for Theoretical Physics (ICTP, Trieste).


\section{Appendix.\ }

In Eq.\ref{longitudinal_pert} the coefficients $M_{r,p+r-q}^{++}$ and $%
L_{r,p+r-q}^{++}$ are a particular case of \ $L_{kp}^{\lambda \tau
},M_{kp}^{\lambda \tau }$\ \ with $\lambda ,\tau =\pm $, defined as: 
\begin{eqnarray}
L_{kp}^{\lambda \tau } &=&\cos \left[ \xi _{k}^{{}}+\lambda \xi
_{p}^{{}}+\tau \left( \zeta _{k}^{{}}+\lambda \zeta _{p}^{{}}\right) \right] 
\notag \\
M_{kp}^{\lambda \tau } &=&\sin \left[ \xi _{k}^{{}}+\lambda \xi
_{p}^{{}}+\tau \left( \zeta _{k}^{{}}+\lambda \zeta _{p}^{{}}\right) \right] 
\end{eqnarray}

In Eq.\ref{transverse_pert} the coefficients $\mathcal{C}_{XY}^{\lambda \tau
}\left( k,q\right) $ with $X,Y=A,B$ are defined as: 
\begin{align}
\mathcal{C}_{XY}^{+-}\left( k,q\right) & =\delta _{XY}\left[
L_{k,k+q}^{++}+\left( 1-2\delta _{XA}\right) M_{k,k+q}^{-+}\right]   \notag
\\
& +\left( 1-\delta _{XY}\right) \left[ L_{k,k+q}^{-+}+\left( 1-2\delta
_{XA}\right) M_{k,k+q}^{++}\right]   \notag \\
\mathcal{C}_{XX}^{-+}\left( k,q\right) & =\mathcal{C}_{YY}^{+-}\left(
k,q\right) \qquad \mathcal{C}_{XY}^{-+}\left( k,q\right) =-\mathcal{C}%
_{YX}^{+-}\left( k,q\right) 
\end{align}

From the longitudinal Kondo term $I_{{}}^{z}$ one obtains the generator $%
R_{{}}^{z}=\sum_{X,Y=A,B}\sum_{m=1,4}R_{mXY}^{z}$\ , Eq.\ref{R_decomposition}%
, as the sum of sixteen contributions. By defining the bosonic operator 
\begin{equation*}
\mathfrak{R}_{mq}=\delta _{m1}a_{p}^{\dagger }a_{q}^{{}}+\delta
_{m2}a_{-q}^{\dagger }a_{-p}^{{}}+\delta _{m3}a_{p}^{\dagger
}a_{-q}^{\dagger }+\delta _{m4}a_{-p}^{{}}a_{q}^{{}}
\end{equation*}%
we have 
\begin{eqnarray}
\mathcal{R}_{mXY}^{z} &=&\sum_{pqr,\sigma }\left[ \delta _{XY}+\left(
1-\delta _{XY}\right) \sigma \right] \left( 1-\delta _{pq}\right) \times  
\notag \\
&&\times \mathcal{X}_{prq}^{mXY}X_{r\sigma }^{\dagger }Y_{p+r-q\sigma }^{{}}%
\mathfrak{R}_{mq}
\end{eqnarray}%
The operators $R_{mXY}^{z}$ depend on the coefficients $\mathcal{X}%
_{pkq\sigma }^{mXY}$. Defining 
\begin{equation*}
F_{kpq}^{XY\lambda \tau }=\mathcal{E}_{k}^{X}-\mathcal{E}_{k+p-q}^{Y}+%
\lambda \hbar \left( \Omega _{q}+\tau \Omega _{p}\right) 
\end{equation*}%
we can write compactly 
\begin{align}
\mathcal{X}_{pkq}^{1XY}& =\delta _{XY}\left( 1-2\delta _{XB}\right) \frac{%
M_{k,p-q+k}^{++}}{F_{kpq}^{XX+-}}\mathrm{Ch}\left( \vartheta _{q}\right) 
\mathrm{Ch}\left( \vartheta _{p}\right)   \notag \\
& -\left( 1-\delta _{XY}\right) \frac{L_{k,p-q+k}^{++}}{F_{kpq}^{XY+-}}%
\mathrm{Ch}\left( \vartheta _{q}\right) \mathrm{Ch}\left( \vartheta
_{p}\right) 
\end{align}%
\begin{align}
\mathcal{X}_{pkq}^{2XY}& =\delta _{XY}\left( 1-2\delta _{XB}\right) \frac{%
M_{k,p-q+k}^{++}}{F_{kpq}^{XX+-}}\mathrm{Sh}\left( \vartheta _{q}\right) 
\mathrm{Sh}\left( \vartheta _{p}\right)   \notag \\
& -\left( 1-\delta _{XY}\right) \frac{L_{k,p-q+k}^{++}}{F_{kpq}^{XY+-}}%
\mathrm{Sh}\left( \vartheta _{q}\right) \mathrm{Sh}\left( \vartheta
_{p}\right) 
\end{align}%
\begin{align}
\mathcal{X}_{pkq}^{3XY}& =\delta _{XY}\left( 1-2\delta _{XB}\right) \frac{%
M_{k,p-q+k}^{++}}{F_{kpq}^{XX++}}\mathrm{Sh}\left( \vartheta _{q}\right) 
\mathrm{Ch}\left( \vartheta _{p}\right)   \notag \\
& -\left( 1-\delta _{XY}\right) \frac{L_{k,p-q+k}^{++}}{F_{kpq}^{XY++}}%
\mathrm{Sh}\left( \vartheta _{q}\right) \mathrm{Ch}\left( \vartheta
_{p}\right) 
\end{align}%
\begin{align}
\mathcal{X}_{pkq}^{4XY}& =\delta _{XY}\left( 1-2\delta _{XB}\right) \frac{%
M_{k,p-q+k}^{++}}{F_{kpq}^{XX-+}}\mathrm{Sh}\left( \vartheta _{q}\right) 
\mathrm{Ch}\left( \vartheta _{p}\right)   \notag \\
& -\left( 1-\delta _{XY}\right) \frac{L_{k,p-q+k}^{++}}{F_{kpq}^{XY-+}}%
\mathrm{Sh}\left( \vartheta _{q}\right) \mathrm{Ch}\left( \vartheta
_{p}\right) 
\end{align}

The generator $R_{{}}^{\perp }$ resulting from the transverse Kondo term $%
I_{{}}^{\perp },$ Eq.\ref{R_decomposition}, has four contributions: $%
\mathcal{R}^{\perp }=\sum_{X,Y=A,B}\mathcal{R}_{XY}^{\perp }$ , namely:%
\begin{eqnarray}
\mathcal{R}_{XY}^{\perp } &=&\frac{J_{K}}{2}\sqrt{\frac{S}{N}}\sum_{kq\sigma
}\left[ \delta _{XY}+\left( 1-\delta _{XY}\right) \sigma \right] \times  
\notag \\
&&\times X_{k\sigma }^{\dagger }Y_{k+q,-\sigma }^{{}}\left( \mathcal{W}%
_{kq}^{XY}a_{q}^{\dagger }+\mathcal{Z}_{kq}^{XY}a_{-q}^{{}}\right) 
\end{eqnarray}%
The coefficients $\mathcal{W}_{kq}^{XY}$ and $\mathcal{Z}_{kq}^{XY}$ are
given by:%
\begin{equation}
\mathcal{W}_{kq}^{XY}=\frac{\left[ \mathrm{Ch}\left( \vartheta _{q}\right) 
\mathcal{C}_{XY}^{+-}\left( k,q\right) +\mathrm{Sh}\left( \vartheta
_{q}\right) \mathcal{C}_{XY}^{-+}\left( k,q\right) \right] }{\left( \mathcal{%
E}_{k}^{X}-\mathcal{E}_{k+q}^{Y}+\hbar \Omega _{q}^{{}}\right) }
\end{equation}%
\begin{equation}
\mathcal{Z}_{kq}^{XY}=\frac{\left[ \mathrm{Sh}\left( \vartheta _{q}\right) 
\mathcal{C}_{XY}^{+-}\left( k,q\right) +\mathrm{Ch}\left( \vartheta
_{q}\right) \mathcal{C}_{XY}^{-+}\left( k,q\right) \right] }{\left( \mathcal{%
E}_{k}^{X}-\mathcal{E}_{k+q}^{Y}-\hbar \Omega _{q}^{{}}\right) }
\end{equation}

We have obtained the effective Hamiltonian in Eq.\ref{H_eff_SW_a}. The
contribution from the perturbations $\frac{1}{2}\left\langle \left[ \mathcal{%
R},I\right] \right\rangle _{fermi}$can be written as the sum of four terms: 
\begin{align}
& \frac{1}{2}\left\langle \left[ \mathcal{R},I\right] \right\rangle _{fermi}
\notag \\
& =\frac{1}{2}\sum_{q}\sum_{X,Y=A,B}\mathcal{T}_{q}^{XY}a_{q}^{\dagger
}a_{q}^{{}}  \notag \\
& +\frac{1}{4}\sum_{q}\sum_{X,Y=A,B}\left( \mathcal{S}_{q}^{XY1}+\mathcal{S}%
_{q}^{XY2}\right) \left( a_{q}^{\dagger }a_{-q}^{\dagger
}+a_{q}^{{}}a_{-q}^{{}}\right)  \notag \\
& +\frac{1}{2}\sum_{q}\hbar \left( \mathcal{D}_{q}^{z+}+\mathcal{D}%
_{q}^{z-}\right) a_{q}^{\dagger }a_{q}^{{}}  \notag \\
& +\frac{1}{4}\sum_{q}\hbar \left( \varpi _{q}^{z+}+\varpi _{q}^{z-}\right)
\left( a_{q}^{\dagger }a_{-q}^{\dagger }+a_{q}^{{}}a_{-q}^{{}}\right)
\end{align}

Assuming the paramagnetic band filling per site $n\leq 1$ so that $%
\left\langle n_{k,\sigma }^{B}\right\rangle =0$ in the ground state, one
finds $\mathcal{T}_{q}^{BB}=\mathcal{S}_{q}^{BB}=0.$

The $\mathcal{T}_{q}^{XY}$ coefficients, by defining $\left\langle N_{k,\pm
q,\sigma }\right\rangle =\left\langle n_{k,-\sigma }^{A}\right\rangle
-\left\langle n_{k\pm q,\sigma }^{A}\right\rangle $ , read:

\begin{align}
\mathcal{T}_{q}^{AA}& =\frac{J_{K}^{2}S}{4N}\sum_{k\sigma }\mathcal{W}%
_{k,q}^{AA}\mathrm{Sh}\left( \vartheta _{q}\right) \mathcal{C}%
_{AA}^{+-}\left( k+q,-q\right) \left\langle N_{k,+q,\sigma }\right\rangle  
\notag \\
& +\frac{J_{K}^{2}S}{4N}\sum_{k\sigma }\mathcal{W}_{k,q}^{AA}\mathrm{Ch}%
\left( \vartheta _{q}\right) \mathcal{C}_{BB}^{+-}\left( k+q,-q\right)
\left\langle N_{k,+q,\sigma }\right\rangle   \notag \\
& +\frac{J_{K}^{2}S}{4N}\sum_{k\sigma }\mathcal{Z}_{k,-q}^{AA}\mathrm{Ch}%
\left( \vartheta _{q}\right) \mathcal{C}_{AA}^{+-}\left( k-q,q\right)
\left\langle N_{k,-q,\sigma }\right\rangle   \notag \\
& +\frac{J_{K}^{2}S}{4N}\sum_{k\sigma }\mathcal{Z}_{k,-q}^{AA}\mathrm{Sh}%
\left( \vartheta _{q}\right) \mathcal{C}_{BB}^{+-}\left( k-q,q\right)
\left\langle N_{k,-q,\sigma }\right\rangle 
\end{align}

\begin{align}
\mathcal{T}_{q}^{AB}& =-\frac{J_{K}^{2}S}{2N}\sum_{k\sigma }\mathcal{W}%
_{k,q}^{AB}\mathrm{Sh}\left( \vartheta _{q}\right) \mathcal{C}%
_{BA}^{+-}\left( k+q,-q\right) \left\langle n_{k,-\sigma }^{A}\right\rangle 
\notag \\
& +\frac{J_{K}^{2}S}{2N}\sum_{k\sigma }\mathcal{W}_{k,q}^{AB}\mathrm{Ch}%
\left( \vartheta _{q}\right) \mathcal{C}_{AB}^{+-}\left( k+q,-q\right)
\left\langle n_{k,-\sigma }^{A}\right\rangle   \notag \\
& -\frac{J_{K}^{2}S}{2N}\sum_{k\sigma }\mathcal{Z}_{k,-q}^{AB}\mathrm{Ch}%
\left( \vartheta _{q}\right) \mathcal{C}_{BA}^{+-}\left( k-q,q\right)
\left\langle n_{k,-\sigma }^{A}\right\rangle   \notag \\
& +\frac{J_{K}^{2}S}{2N}\sum_{k\sigma }\mathcal{Z}_{k,-q}^{AB}\mathrm{Sh}%
\left( \vartheta _{q}\right) \mathcal{C}_{AB}^{+-}\left( k-q,q\right)
\left\langle n_{k,-\sigma }^{A}\right\rangle 
\end{align}%
and 
\begin{eqnarray}
\mathcal{T}_{q}^{BA} &=&\frac{J_{K}^{2}S}{2N}\sum_{k}\mathcal{W}_{k,q}^{BA}%
\mathrm{Sh}\left( \vartheta _{q}\right) \mathcal{C}_{AB}^{+-}\left(
k+q,-q\right) \left\langle n_{k+q,\sigma }^{A}\right\rangle   \notag \\
&&-\frac{J_{K}^{2}S}{2N}\sum_{k}\mathcal{W}_{k,q}^{BA}\mathrm{Ch}\left(
\vartheta _{q}\right) \mathcal{C}_{BA}^{+-}\left( k+q,-q\right) \left\langle
n_{k+q,\sigma }^{A}\right\rangle   \notag \\
&&+\frac{J_{K}^{2}S}{2N}\sum_{k}\mathcal{Z}_{k,-q}^{BA}\mathrm{Ch}\left(
\vartheta _{q}\right) \mathcal{C}_{AB}^{+-}\left( k-q,q\right) \left\langle
n_{k-q,\sigma }^{A}\right\rangle   \notag \\
&&-\frac{J_{K}^{2}S}{2N}\sum_{k}\mathcal{Z}_{k,-q}^{BA}\mathrm{Sh}\left(
\vartheta _{q}\right) \mathcal{C}_{BA}^{+-}\left( k-q,q\right) \left\langle
n_{k-q,\sigma }^{A}\right\rangle   \notag \\
&&
\end{eqnarray}

The $\mathcal{S}_{q}^{XY1}$ coefficients read:%
\begin{eqnarray}
\mathcal{S}_{q}^{AA1} &=&\frac{J_{K}^{2}S}{2N}\sum_{k}\mathcal{W}_{k,-q}^{AA}%
\mathrm{Ch}\left( \vartheta _{q}\right) \mathcal{C}_{AA}^{+-}\left(
k-q,q\right) \left\langle N_{k,-q,\sigma }\right\rangle   \notag \\
&&+\frac{J_{K}^{2}S}{2N}\sum_{k}\mathcal{W}_{k,-q}^{AA}\mathrm{Sh}\left(
\vartheta _{q}\right) \mathcal{C}_{BB}^{+-}\left( k-q,q\right) \left\langle
N_{k,-q,\sigma }\right\rangle   \notag \\
&&
\end{eqnarray}%
\begin{eqnarray}
\mathcal{S}_{q}^{AB1} &=&-\frac{J_{K}^{2}S}{2N}\sum_{k}\mathcal{W}%
_{k,-q}^{AB}\mathrm{Ch}\left( \vartheta _{q}\right) \mathcal{C}%
_{BA}^{+-}\left( k-q,q\right) \left\langle n_{k,-\sigma }^{A}\right\rangle  
\notag \\
&&+\frac{J_{K}^{2}S}{2N}\sum_{k}\mathcal{W}_{k,-q}^{AB}\mathrm{Sh}\left(
\vartheta _{q}\right) \mathcal{C}_{AB}^{+-}\left( k-q,q\right) \left\langle
n_{k,-\sigma }^{A}\right\rangle   \notag \\
&&
\end{eqnarray}%
\begin{eqnarray}
\mathcal{S}_{q}^{BA1} &=&\frac{J_{K}^{2}S}{2N}\sum_{k}\mathcal{W}_{k,-q}^{BA}%
\mathrm{Ch}\left( \vartheta _{q}\right) \mathcal{C}_{AB}^{+-}\left(
k-q,q\right) \left\langle n_{k-q,\sigma }^{A}\right\rangle   \notag \\
&&-\frac{J_{K}^{2}S}{2N}\sum_{k}\mathcal{W}_{k,-q}^{BA}\mathrm{Sh}\left(
\vartheta _{q}\right) \mathcal{C}_{BA}^{+-}\left( k-q,q\right) \left\langle
n_{k-q,\sigma }^{A}\right\rangle   \notag \\
&&
\end{eqnarray}

The coefficients $\mathcal{S}_{q}^{XY2}$ can be obtained from $\mathcal{S}%
_{q}^{XY1}$ by interchanging $\mathcal{W}_{k,-q}^{XY}$ and $\mathrm{Ch}%
\left( \vartheta _{q}\right) $ respectively with $\mathcal{Z}_{k,-q}^{XY}$
and $\mathrm{Sh}\left( \vartheta _{q}\right) .$

To write down the coefficients $\mathcal{D}_{q}^{z\pm }$ and $\varpi
_{q}^{z\pm }$ of Eq.\ref{H_eff_SW_b} it is convenient to introduce:%
\begin{equation*}
\mathrm{Ch}\left( \vartheta _{q}+\vartheta _{p}\right) =\mathfrak{C}%
_{qp}\quad \mathrm{Sh}\left( \vartheta _{q}+\vartheta _{p}\right) =\mathfrak{%
S}_{qp}
\end{equation*}%
By defining  
\begin{eqnarray}
\mathcal{L}_{pqr}^{+} &=&M_{r,p-q+r}^{++}\left( \mathcal{X}%
_{q,p-q+r,p}^{AA1}+\mathcal{X}_{-p,p-q+r,-q}^{AA2}\right) \mathfrak{C}%
_{qp}\left\langle n_{p-q+r,\sigma }^{A}\right\rangle   \notag \\
&&-L_{r,p-q+r}^{++}\left( \mathcal{X}_{q,p-q+r,p}^{AB1}+\mathcal{X}%
_{-p,p-q+r,-q}^{AB2}\right) \mathfrak{C}_{qp}\left\langle n_{p-q+r,\sigma
}^{A}\right\rangle   \notag \\
&&-M_{r,p-q+r}^{++}\left( \mathcal{X}_{q,p-q+r,p}^{AA3}+\mathcal{X}%
_{-p,p-q+r,-q}^{AA3}\right) \mathfrak{S}_{qp}\left\langle n_{r\sigma
}^{A}\right\rangle   \notag \\
&&+L_{r,p-q+r}^{++}\left( \mathcal{X}_{q,p-q+r,p}^{BA3}+\mathcal{X}%
_{-p,p-q+r,-q}^{BA3}\right) \mathfrak{S}_{qp}\left\langle n_{r\sigma
}^{A}\right\rangle   \notag \\
&&
\end{eqnarray}%
and 
\begin{eqnarray}
\mathcal{L}_{pqr}^{-} &=&-M_{r,p-q+r}^{++}\left( \mathcal{X}%
_{-p,p-q+r,-q}^{AA1}+\widetilde{\mathcal{X}}_{q,p-q+r,p}^{AA2}\right) 
\mathfrak{C}_{qp}\left\langle n_{r\sigma }^{A}\right\rangle   \notag \\
&&+L_{r,p-q+r}^{++}\left( \mathcal{X}_{-p,p-q+r,-q}^{BA1}+\mathcal{X}%
_{q,p-q+r,p}^{BA2}\right) \mathfrak{C}_{qp}\left\langle n_{r\sigma
}^{A}\right\rangle   \notag \\
&&+M_{r,p-q+r}^{++}\left( \mathcal{X}_{q,p-q+r,p}^{AA4}+\mathcal{X}%
_{-p,p-q+r,-q}^{AA4}\right) \mathfrak{S}_{qp}\left\langle n_{p-q+r,\sigma
}^{A}\right\rangle   \notag \\
&&-L_{r,p-q+r}^{++}\left( \mathcal{X}_{q,p-q+r,p}^{AB4}+\mathcal{X}%
_{-p,p-q+r,-q}^{AB4}\right) \mathfrak{S}_{qp}\left\langle n_{p-q+r,\sigma
}^{A}\right\rangle   \notag \\
&&
\end{eqnarray}%
we can write 
\begin{equation}
\hbar \mathcal{D}_{q}^{z\pm }=\frac{J_{K}^{2}}{2}\left( \frac{2}{N}\right)
^{2}\sum_{pr}\left( 1-\delta _{pq}\right) \mathcal{L}_{pqr}^{\pm }
\end{equation}

Next, by defining%
\begin{eqnarray}
\mathcal{G}_{pqr}^{+} &=&M_{r,p-q+r}^{++}\left( \mathcal{X}%
_{q,p-q+r,p}^{AA1}+\mathcal{X}_{-p,p-q+r,-q}^{AA2}\right) \mathfrak{S}%
_{qp}\left\langle n_{p-q+r,\sigma }^{A}\right\rangle   \notag \\
&&-L_{r,p-q+r}^{++}\left( \mathcal{X}_{q,p-q+r,p}^{AB1}+\mathcal{X}%
_{-p,p-q+r,-q}^{AB2}\right) \mathfrak{S}_{qp}\left\langle n_{p-q+r,\sigma
}^{A}\right\rangle   \notag \\
&&-M_{r,p-q+r}^{++}\left( \mathcal{X}_{q,p-q+r,p}^{AA3}+\mathcal{X}%
_{-p,p-q+r,-q}^{AA3}\right) \mathfrak{C}_{qp}\left\langle n_{r\sigma
}^{A}\right\rangle   \notag \\
&&+L_{r,p-q+r}^{++}\left( \mathcal{X}_{q,p-q+r,p}^{BA3}+\mathcal{X}%
_{-p,p-q+r,-q}^{BA3}\right) \mathfrak{C}_{qp}\left\langle n_{r\sigma
}^{A}\right\rangle   \notag \\
&&
\end{eqnarray}

and 
\begin{eqnarray}
\mathcal{G}_{pqr}^{-} &=&-M_{r,p-q+r}^{++}\left( \mathcal{X}%
_{-p,p-q+r.-q}^{AA1}+\mathcal{X}_{q,p-q+r,p}^{AA2}\right) \mathfrak{S}%
_{qp}\left\langle n_{r\sigma }^{A}\right\rangle   \notag \\
&&+L_{r,p-q+r}^{++}\left( \mathcal{X}_{-p,p-q+r,-q}^{BA1}+\mathcal{X}%
_{q,p-q+r,p}^{BA2}\right) \mathfrak{S}_{qp}\left\langle n_{r\sigma
}^{A}\right\rangle   \notag \\
&&+M_{r,p-q+r}^{++}\left( \mathcal{X}_{q,p-q+r,p}^{AA4}+\mathcal{X}%
_{-p,p-q+r,-q}^{AA4}\right) \mathfrak{C}_{qp}\left\langle n_{p-q+r,\sigma
}^{A}\right\rangle   \notag \\
&&-L_{r,p-q+r}^{++}\left( \mathcal{X}_{q,p-q+r,p}^{AB4}+\mathcal{X}%
_{-p,p-q+r,-q}^{AB4}\right) \mathfrak{C}_{qp}\left\langle n_{p-q+r,\sigma
}^{A}\right\rangle   \notag \\
&&
\end{eqnarray}

we can write 
\begin{equation}
\hbar \mathcal{\varpi }_{q}^{z\pm }=\frac{J_{K}^{2}}{2}\left( \frac{2}{N}%
\right) ^{2}\sum_{pr}\left( 1-\delta _{pq}\right) \mathcal{G}_{pqr}^{\pm }
\end{equation}

Using this set of contributions, the coefficients $\Theta _{q}$ and $\Psi
_{q}$ in Eq.\ref{H_eff_SW_b} are defined as: \ \ 
\begin{equation}
\hbar \Theta _{q}=\frac{1}{2}\sum_{X,Y=A,B}\mathcal{T}_{q}^{XY}+\frac{1}{2}%
\left( \hbar \mathcal{D}_{q}^{z+}+\hbar \mathcal{D}_{q}^{z-}\right) 
\label{Phi_q}
\end{equation}%
\begin{equation}
\hbar \Psi _{q}=\sum_{X,Y=A,B}\left( \frac{\mathcal{S}_{q}^{XY1}+\mathcal{S}%
_{q}^{XY2}}{4}\right) +\hbar \left( \frac{\varpi _{q}^{z+}+\varpi _{q}^{z-}}{%
4}\right)   \label{Psi_q}
\end{equation}




%
%

%
%
%
%

\newpage 

\mediumtext
\begin{table}
\caption{Conduction band AF spin polarization $<s>$  dependence 
on model parameters, for the cases 
depicted  in respective Figs. 3, 4 and 5. 
 In first column we enter the relative  
Kondo coupling magnitude (cases of Fig. 3: notice that $ J_K/W 
\sim  J_K/J_H \times 10^{-4}$, here), while  
in second column is the corresponding band polarization $<s>$ obtained. 
Similarly in  third column we enter the electron 
correlation $U/W$ cases from Fig. 4, and next  
the corresponding  $<s>$. In 5th column we enter filling $n$ values 
of Fig. 5, and  in last column the corresponding  
$<s>$ values.
} \label{table1}
\begin{tabular}{cccccc}
\multicolumn{1}{c}{$ J_K/J_H$}
&\multicolumn{1}{c}{$<s>$}&
\multicolumn{1}{c}{$\qquad U/W$} &\multicolumn{1}{c}{$<s>$}
&\multicolumn{1}{c}{ $\qquad n$} &\multicolumn{1}{c}{$<s>$} \\
\tableline
 0.1 & 0.2152 & \qquad 0.008 & 0.0002 & \qquad 0.999 & 0.2153 \\
 1.  & 0.2153 & \qquad 0.08 & 0.001   &  \qquad 0.8   & 0.1675 \\
 5.  & 0.2157 & \qquad 0.25 & 0.215   &  \qquad 0.7   & 0.1329  \\
 10.  & 0.2163 & \qquad 0.5  & 0.406   & \qquad 0.6   & 0.0796  \\
 15.  & 0.2168 & \qquad 0.83 & 0.480   & \qquad 0.4   & 0.00006 \\

\end{tabular}
\end{table}

\end{document}